\documentclass[preprint,12pt]{elsarticle}

\usepackage{amssymb}
\usepackage{multirow}
\usepackage{booktabs}
\usepackage{amssymb}
\usepackage{mathrsfs}
\usepackage{longtable}
\usepackage{lscape}
\usepackage{tabularx}
\usepackage{epsfig}
\usepackage{colortbl}
\usepackage{subfigure}
\journal{Physica A}

\begin{document}

\begin{frontmatter}



\title{A new structure entropy of complex networks based on Tsallis nonextensive statistical mechanics}


\author[swu]{Qi Zhang}
\author[swu,hh]{Xi Lu}
\author[swu]{Meizhu Li}

\author[swu,NWPU]{Yong Deng\corref{cor}}
\ead{ydeng@swu.edu.cn, prof.deng@hotmail.com}
\author[vu]{Sankaran Mahadevan}

\cortext[cor]{Corresponding author: Yong Deng, School of Computer and Information Science, Southwest University, Chongqing, 400715, China.}

\address[swu]{School of Computer and Information Science, Southwest University, Chongqing, 400715, China}

\address[hh]{School of Hanhong, Southwest University, Chongqing, 400715, China}

\address[NWPU]{School of Automation, Northwestern Polytechnical University, Xian, Shaanxi 710072, China}

\address[vu]{School of Engineering, Vanderbilt University, Nashville, TN, 37235, USA}

\begin{abstract}
    The structure entropy is one of the most important parameters to describe the structure property of the complex networks. Most of the existing structure entropies are based on the degree or the betweenness centrality. In order to describe the structure property of the complex networks more reasonably, a new structure entropy of the complex networks based on the Tsallis nonextensive statistical mechanics is proposed in this paper. The influence of the degree and the betweenness centrality on the structure property is combined in the proposed structure entropy.
    Compared with the existing structure entropy, the proposed structure entropy is more reasonable to describe the structure property of the complex networks in some situations.
\end{abstract}
\end{frontmatter}

\section{Introduction}
The complex network is composed of many interacting parts \cite{kim2008complex,newman2003structure}. In the real world, many systems can be modelled as a complex network. Many of the researches on the complex networks are based on the structure property, such as the "Small-world network"\cite{watts1998collective}, the "Scale-free network"\cite{barabasi2009scale,cohen2003scale} and so on.

Since the structure property of the complex networks play an very important role, it is heavily studied \cite{xiao2008symmetry,xu2013degree,tan2004network}. Many applications of the complex networks based on the structure entropy are published \cite{wang2006entropy,curado1991generalized,crucitti2004error}.

The structure entropy is one of the most important methods to describe the structure property of the complex networks, such as the degree structure entropy \cite{anand2009entropy} and the betweenness structure entropy \cite{ZhangLD14}.
In the degree structure entropy, the basic factor is the degree of the nodes. The degree of the nodes is a local structure property of the complex network.
Therefore, the global structure property of the complex network is not illuminated in the degree structure entropy.
In the betweenness structure entropy, the basic factor is based on the betweenness of the nodes.
The betweenness of the node is a global characteristic of the complex networks, which is only based on the shortest path in the network. Therefore, the details of the local structure property in the complex networks are ignored in the betweenness structure entropy.
It seems necessary that both of the degree structure entropy and the betweenness structure entropy should be improved.

In order to describe the structure property of the complex networks comprehensively, in this paper, a new structure entropy of the complex networks is proposed based on the Tsallis nonextensive statistical mechanics \cite{tsallis1988possible,tsallis1994nonextensive,tsallis1996anomalous}. The influence of the degree and the betweenness centrality on the structure property is combined in the proposed structure entropy. Compared with the existing structure entropy, the proposed structure entropy is more reasonable to describe the structure property of the complex networks in some situations.

The rest of this paper is organised as follows. Section \ref{Rreparatorywork} introduces some preliminaries of this work, such as the degree distribution of the complex network, the betweenness centrality of the complex network, the Tsallis nonextensive statistical mechanics and the existing structure entropy. In section \ref{new}, a new structure entropy of the complex networks based on the Tsallis nonextensive statistical mechanics is proposed. The application of the proposed method is illustrated in section \ref{application}. Conclusion is given in Section \ref{conclusion}.

\section{Preliminaries}
\label{Rreparatorywork}

The structure entropy is a global characteristic which can be used to describe the structure complexity of the complex networks. The smaller the value of the structure entropy, the more orderly the complex networks.
The introduction of the concepts which will be used in the new structure entropy are shown as follows.
\subsection{Degree distribution}
\label{Degree}
The degree of one node in a network is the number of the edges connected to the node. Most of the properties of the complex network are based on the degree distribution, such as the clustering coefficient, the community structure and so on. In the network, ${Degree(i)}$ represents the degree of the $i$th node. To calculate the structure entropy of the complex networks, the degree of the complex networks is defined as follows \cite{xiao2008symmetry,xu2013degree}:

\begin{equation}\label{KI}
{p_i} = \frac{{{Degree(i)}}}{{\sum\limits_{i = 1}^n {{Degree(i)}} }}
\end{equation}

Where the ${p_i}$ represents the new degree which will be used in the structure entropy \cite{newman2003structure}.

\subsection{Betweenness centrality}
\label{Betweenness}

The betweenness centrality of the complex networks is another property of the nodes. It is defined based on the shortest-path in the network \cite{barthelemy2004betweenness}. The details of the definition of the betweenness are shown as follows \cite{barthelemy2004betweenness}.

\begin{equation}\label{Betweenness}
{\upsilon(i)}= \sum\limits_{s \ne v \ne t} {\frac{{{\sigma _{st}}\left( \nu  \right)}}{{{\sigma _{st}}}}}
\end{equation}

Where the ${\upsilon(i)}$ represents the betweenness of the ${i}$th node. The ${{{\sigma _{st}}}}$ represents the total number of the shortest paths from node $s$ to node $t$. The ${{\sigma _{st}}\left( i  \right)}$ represents the number of the shortest paths that pass through the ${i}$th node.

\subsection{Existing structure entropy}
\label{Structure entropy}
The existing structure entropy of the complex networks is based on the information entropy \cite{shannon2001mathematical}, which is defined by Shannon as follows \cite{shannon2001mathematical}.

\begin{equation}\label{Shannon-entropy}
 {E_{Shannon}}{\rm{ = }}\sum\limits_{i = 1}^n {{p_i}} \log ({p_i})
\end{equation}

The degree distribution is widely used in most of the existing structure entropies of complex networks, such as the structure entropy proposed by Xu et.al \cite{xu2013degree}. It is defined as follows\cite{wang2006entropy}.

\begin{equation}\label{degree-entropy}
E_{deg} =  - k\sum\limits_{i = 1}^N {{p_i}\log {p_i}}
\end{equation}

Where ${p_i}$ is defined as follows:
\begin{equation}\label{pj}
{{p_i}} = {\textstyle{{Degree(i)} \over {\sum\limits_{i = 1}^N {Degree(i)} }}}
\end{equation}

The $Degree(j)$ in the Eq.(\ref{pj}) represents the degree of the $j$th node and the $N$ represents the total number of the nodes in the network.

Recently, a betweenness structure entropy of complex networks is proposed in \cite{ZhangLD14}. It is defined as follows.

\begin{equation}\label{betweenness-entropy}
{E_{bet}} =  - \sum\limits_{i = 1}^n {{{p_i}'}\log {{p_i}'}}
\end{equation}

Where the $p_i$ is defined as follows:

\begin{equation}\label{pi}
{{p_i}'} = \frac{{\upsilon (i)}}{{\sum\limits_{i = 1}^n {\upsilon (i)} }}
\end{equation}

The $\upsilon (i)$ in the Eq.(\ref{pi}) represents the betweenness which is defined in section \ref{Betweenness}.

However, both of the two kinds of structure entropies are not comprehensive. The degree structure entropy ignores the global property of the complex networks while the betweenness structure ignores the local property of the complex networks. A new structure entropy is needed to describe the structure property of the complex networks more comprehensively.

\subsection{Tsallis nonextensive statistical mechanics}
\label{Tsallis entropy}
The entropy is defined by Clausius for thermodynamics \cite{clausius1867mechanical}. For a finite discrete set of probabilities the definition of the Boltzmann-Gibbs entropy \cite{gibbs2010elementary} is given as follows:

\begin{equation}\label{S_BG}
{S_{BG}} =  -k \sum\limits_{i = 1}^N {{p_i}} \ln {p_i}
\end{equation}

The conventional constant $k$ is the Boltzmann universal constant for thermodynamic systems. The value of $k$ will be taken to be unity in information theory \cite{tsallis2010nonadditive,shannon2001mathematical}.

In 1988, a more general form for entropy have been proposed by Tsallis  \cite{tsallis1988possible}. It is shown as follows:

\begin{equation}\label{S_q}
{S_q} =  -k \sum\limits_{i = 1}^N {{p_i}} {\ln _q}{p_i}
\end{equation}

The $q-logarithmic$ function in the Eq. (\ref{S_q}) is presented as follows \cite{tsallis2010nonadditive,tsallis1998role,tsallis1996generalized}:
\begin{equation}\label{ln_q}
{\ln _q}{p_i} = \frac{{{p_i}^{1 - q} - 1}}{{1 - q}}({p_i} > 0;q \in \Re ;l{n_1}{p_i} = ln{p_i})
\end{equation}

Based on the Eq. (\ref{ln_q}), the Eq. (\ref{S_q}) can be rewritten as follows:
\begin{equation}\label{S_q1}
{S_q} = k\frac{{1 - \sum\limits_{i = 1}^N {{p_i}^q} }}{{q - 1}}
\end{equation}

Where ${N}$ is the number of the subsystems.

Based on the Tsallis entropy, the nonextensive theory is established by Tsallis et.al \cite{tsallis2010nonadditive}.

In the nonextensive statistical mechanics, the relationship between the subsystem and the system can be described as the nonevtensive additivity.
The nonextensive parameter ${q}$ in the nonextensive statistical mechanics is used to describe the nonextensive additivity of the system.
When the value of the nonextensive parameter ${q}$ is equal to 1, the nonextensive additivity in the system is degenerated to the classical additivity \cite{tsallis2010nonadditive,tsallis2014introduction}.

According to the nonextensive, the relationship between the basic factor and the structure entropy can be described as a nonextensive additivity.

\section{A new structure entropy of the complex networks}
\label{new}

In Tsallis nonextensive statistical mechanics, the nonextensive parameter $q$ is used to describe the nonextensive additivity in the system \cite{tsallis2001nonextensiveLNP,tsallis2002entropic,tsallis2004nonextensive,tsallis2009introduction}.
Based on the principle of the nonextensive statistical mechanics, the betweenness is used as a nonextensive parameter to modify the construction of the proposed structure entropy.

In the Tsallis nonextensive statistic mechanics, each system has a nonextensive parameter ${q}$ to describe the nonextensive additivity among it. In order to describe the nonextensive additivity of the structure entropy in details, in the proposed structure entropy, each structure entropy has a series of nonextensive parameters ${q_i}$ to describe the nonextensice additivity in it.
In other words, each basic factor has an nonextensive parameter to describe the nonextensive additivity between it and the structure entropy \cite{tsallis2013black,tsallis2014introduction}.

In the proposed structure entropy, the nonextensive additivity is used to describe the relationship between the basic factor and the structure entropy. The betweenness of each nodes is used as the nonextensive parameter to describe the nonextensive additivity. It means that, the influences of degree and betweenness on the structure property are combined in the proposed structure entropy by the nonextensive additivity \cite{tsallis2014news,tsallis2013nonlinear,tsallisPhysRevE.89.052135,Ruiz2013491,Mariz20123088}. Therefore, the proposed structure entropy is a generalised structure entropy of the complex network. The details of the definition of the proposed structure entropy are shown as follows:

\begin{equation}\label{New_structure_entropy}
  {E_{T}} = \sum\limits_{i = 1}^n {{h_i}} \log ({h_i})
\end{equation}

Where the ${E_{T}}$ represents the new structure entropy. The ${h_i}$ is the basic factor in the new structure entropy, which is defined based on the degree, the betweenness centrality and the Tsallis entropy \cite{tsallis2009nonadditiveBJP,tsallis2009nonadditiveEPJA,tsallis2010nonadditive}. The details of ${h_i}$ is shown as follows.

\begin{equation}\label{h_i}
{h_i} = \frac{{{p_i}^{{q_i}}}}{{\sum\limits_{i = 1}^n {{p_i}^{{q_i}}} }}
\end{equation}

Where the ${p_i}$ is the degree of the ${i}$th node defined in the Eq.(\ref{pj}). The ${q_i}$ represents the nonextensive parameter which is defined based on the betweenness. The nonextensive parameter ${q_i}$ is defined as follows:
\begin{equation}\label{q_i}
{q_i} = 1 + (\upsilon (max) - \upsilon (i)), \left\{ {\upsilon (max) = max\left[ {\upsilon (i),(i = 1,2,3, \cdots ,n)} \right]} \right\} \end{equation}

According to the Eq.(\ref{h_i}) and Eq.(\ref{q_i}) the definition of the new structure entropy can be represented as follows.

\begin{equation}\label{E_T_2}
{E_T} = \sum\limits_{i = 1}^n {\frac{{{p_i}^{{q_i}}}}{{\sum\limits_{i = 1}^n {{p_i}^{{q_i}}} }}} \log (\frac{{{p_i}^{{q_i}}}}{{\sum\limits_{i = 1}^n {{p_i}^{{q_i}}} }})
\end{equation}


In the definition of the structure entropy, the value of ${p_i}$ is smaller than 1. The value of ${q_i}$ is bigger than 1.
In other words, the nonextensive parameter ${q_i}$ is used to decline the influence of degree on the structure entropy. Therefore, in the new structure entropy, the influence of degree on the structure entropy is smaller than that in the existing structure entropy.

\begin{figure}[htbp]
  \centering
  \subfigure[Network A]{
    \label{example:subfig:a} 
    \includegraphics[scale=3.5]{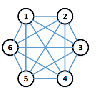}}
    \subfigure[Network B]{
    \label{example:subfig:b} 
    \includegraphics[scale=3.5]{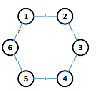}}
    \subfigure[Network C]{
    \label{example:subfig:c} 
    \includegraphics[scale=3.5]{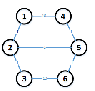}}
      \subfigure[Network D]{
    \label{example:subfig:d} 
    \includegraphics[scale=3.5]{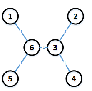}}
  \caption{All of the networks have 6 nodes with different structure. The Network A is a global coupling network. The Network B is a simple nearest-neighbor coupled network. The Network C is a symmetrical network. The Network D is a dumbbell network. The weight of the edges in the Network A and Network D is equal to 1.  The weight of each edges in Network B and Network C is different. The four networks have special structure property.}\label{model}
\end{figure}

To prove the rationality of the proposed structure entropy, 4 networks with different structure are established. The details of the 4 networks are shown in Fig. \ref{model}.

The structure entropy of the 4 networks is calculated with the degree structure entropy, the betweenness structure entropy and the proposed structure entropy. The results are shown in Table \ref{test-networks}.

\begin{table}[htbp]
  \centering
  \caption{The structure entropy of the test networks}
    \begin{tabular}{ccccc}
  \hline
   & Network A & Network B & Network C & Network D \\
   \hline
  $E_{deg}$ & \textbf{1.7918} & \textbf{1.7918} & 1.7721 & 1.6434 \\
  $E_{bet}$ & 1.7918 & 1.5955 & \textbf{1.5763} & \textbf{1.5763} \\
  $E_{T}$ & 1.7918 & 1.7808 &1.6932 & 1.4607 \\
  \hline
    \end{tabular}%
  \label{test-networks}%
\end{table}%
The value of the structure entropy represents the degree of the complexity of the networks. Based on different values of structure entropy, the degree of the complexity in the networks can be ordered. The details are shown in the Table \ref{result-networks}.

\begin{table}[htbp]
  \centering
  \caption{The degree of the complexity in those test networks}
    \begin{tabular}{cccccccc}
   \hline
  $E_{deg}$ & Network A $\equiv$& Network B $>$& Network C$>$ & Network D \\
  $E_{bet}$ & Network A $>$ & Network B $>$& Network C$\equiv$ & Network D \\
  $E_{T}$ & Network A $>$ & Network B $>$& Network C$>$ & Network D \\
  \hline
    \end{tabular}%
  \label{result-networks}%
\end{table}%

The value of degree structure entropy for Network A is equal to the value of Network B. It means that under the degree structure entropy, the two networks have the same degree of structure complexity. However, it is clear shown in the Fig. \ref{model} that the two networks have different structure and structure complexity.
According to the betweenness structure entropy, the value of the structure entropy of Network C and Network D is equal to each other. It means that depends on the betweenness structure entropy, the two networks have the same structure complexity too. In other words, for the networks which have a special structure, the existing structure entropy can not identify the difference among them.
However, the value of the proposed structure entropy for the four networks are different from each other. It means that the proposed structure entropy can be used to identify the difference of the structure in the networks.
Compared with the existing structure entropy, the proposed structure entropy is more reasonable to describe the structure property of the complex network, especially for those networks which have special structure.
\section{Application}
\label{application}
\begin{figure}[Htbp]
  \centering
  \subfigure[The SF-network A]{
    \label{subfig:a} 
    \includegraphics[scale=0.8]{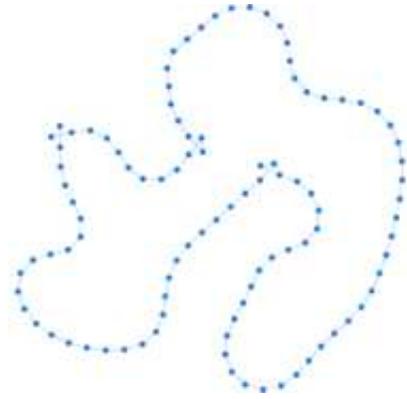}}
    \subfigure[The SF-network B]{
    \label{subfig:b} 
    \includegraphics[scale=0.8]{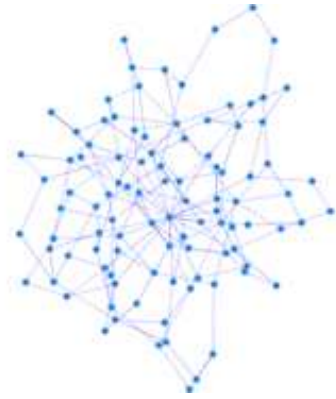}}
    \subfigure[The SF-network C]{
    \label{subfig:c} 
    \includegraphics[scale=0.8]{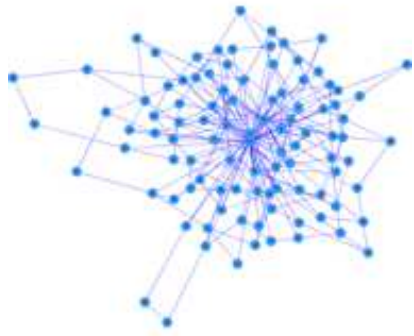}}
      \subfigure[The SF-network D]{
    \label{subfig:d} 
    \includegraphics[scale=0.8]{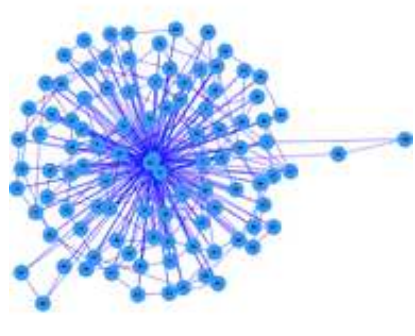}}
  \caption{This figure is used to show of the Scale-Free networks which is establishes by us. First, we build a nearest-neighbor coupled network with 100 nodes. Second, each node in the network has a chance to connect to other nodes in randomly. Then selecting those nodes which have bigger value of the degree as a set of the central node. Third, each node has a probability to connect to those central nodes. The probability is based on the degree of those central nodes. Then repeating the step 2 and step 3 in the times $K$. The SF-network A is the nearest-neighbor coupled network. The SF-network B is Scale-Free networks with the times $K=3$. The times of the SF-network C $K=10$. The times of the SF-networks D is equal to 50. It is clear that following the increase of the times $K$, the $Scale-Free$ property is increased too.}\label{SFmodel}
\end{figure}

To prove the efficiency of the proposed structure entropy, we establish a series Scale-Free networks. The networks are growing from the simple nearest-neighbor coupled network. The process of the growing of the Scale-Free network is shown in the Fig. \ref{SFmodel}. The proposed structure entropy is used to calculate the structure entropy of the network, among the process of the growing of the Scale-Free network. The results are shown in Table \ref{tab:addlabel}.

\begin{table}[htbp]
  \centering
  \caption{The structure entropy of the Scale-Free networks with different times of $K$}
   \begin{tabular}{cccccc}
     \hline
       Network  & 1 & 2 & 3 &  4 &  5 \\
      \hline
    Nodes  & 100   & 300   & 500   & 700   & 1000 \\
      \hline
     $E_{{T}(K=1)}$     & 4.3062 & 6.0212 & 5.4723 & 6.3822 & 6.7634 \\
     $E_{{T}(K=5)}$     & 3.0716 & 5.0076 & 4.4328 & 5.2377 & 5.5498 \\
     $E_{{T}(K=10)}$     & 3.0962 & 4.8086 & 4.3464 & 5.1341 & 5.5015 \\
     $E_{{T}(K=15)}$     & 3.0476 & 4.7993 & 4.2435 & 4.9345 & 5.3872 \\
     $E_{{T}(K=20)}$     & 2.9768 & 4.6164 & 4.2687 & 5.0449 & 5.366 \\
     $E_{{T}(K=25)}$     & 2.9670 & 4.6280 & 4.2488 & 5.0093 & 5.2889 \\
     $E_{{T}(K=30)}$     & 2.9605 & 4.5793 & 4.0394 & 4.8781 & 5.2310 \\
     $E_{{T}(K=50)}$     & 2.9548 & 4.4043 & 4.0477 & 4.6107 & 5.1613 \\
      \hline
    \end{tabular}%
  \label{tab:addlabel}%
\end{table}%

Table \ref{tab:addlabel} shows that with the increase of the times $K$, the value of the Tsallis structure entropy is reduced. In the Table \ref{tab:addlabel}, the value of ${E_{T}}$ of the Scale-Free networks with $K=50$ is small than others. The larger the value of ${K}$, the smaller the value of ${E_{T}}$.
The results of our test is coincided to the regular of the Scale-Free networks. It means that the proposed structure entropy can be used to describe the structure complexity of the complex network.

To find the difference between the existing structure entropy and the proposed structure entropy in this paper, the existing structure entropy and the proposed structure entropy are used to calculate the structure entropy of the real networks, such as the  US-airport network \cite{networkdata}, Email networks \cite{networkdata}, the Germany highway networks \cite{nettt},the Zachary's Karate Club network \cite{uci} and the protein-protein interaction network in budding yeast \cite{networkdata}. The results are shown in Table \ref{real-networks}.

\begin{table}[htbp]
  \centering
  \caption{The structure entropy of the real networks}
    \begin{tabular}{llllll}
    \hline
    Network & Nodes  & Edges & $E_{deg}$ & $E_{bet}$ &$E_T$\\
    \hline
    US-Airport \cite{networkdata} & 500   & 5962  & 5.025 & 3.3512 &  1.1016 \\
    Email \cite{networkdata} & 1133  & 10902 & 6.6310 & 5.3876 & 1.5236\\
    Yeast \cite{networkdata}  & 2375  & 23386 & 7.0539 & 6.1644 & 2.1900\\
    Germany highway \cite{nettt}& 1168  & 2486  & 6.9947 & 5.6383  &1.9499\\
    \hline
    \end{tabular}%
  \label{real-networks}%
\end{table}%

In the Table  \ref{real-networks}, the value of the ${E_T}$ is smaller than others. It means that based on the new structure entropy, those real networks seems more orderly.

According to the value of the networks, the most orderly network in the those four networks is the Us Airport \cite{networkdata}. The details of the networks complexity are shown in Table \ref{result-real-networks}.

\begin{table}[htbp]
\small
\addtolength{\tabcolsep}{-6pt}
  \centering
  \caption{The degree of the complexity in those real networks}
    \begin{tabular}{cc}
   \hline
  $E_{deg}$ & Yeast \cite{networkdata} $>$ Germany highway \cite{nettt} $>$ Email \cite{networkdata}$>$ US-Airport \cite{networkdata} \\
  $E_{bet}$ & Yeast \cite{networkdata} $>$ Germany highway \cite{nettt} $>$ Email \cite{networkdata}$>$ US-Airport \cite{networkdata} \\
  $E_{T}$ & Yeast \cite{networkdata} $>$ Germany highway \cite{nettt}$>$ Email \cite{networkdata}$>$ US-Airport \cite{networkdata} \\
  \hline
    \end{tabular}%
  \label{result-real-networks}%
\end{table}%
The results shows that, depends on the proposed structure entropy and the existing structure entropy, those networks have the same order in the degree of complexity. It means that the proposed structure entropy is effective to describe the structure property of the complex networks.

\section{Conclusion}
\label{conclusion}
In the complex networks, the structure entropy can be used to describe the structure property from the global view. It is an important structure characteristic. Most of the existing structure entropies are based on one of the structure property, such as the degree structure entropy is based on the degree distribution and the betweenness structure entropy is based on the betweenness centrality. In this paper, the influences of the degree and betweenness on the structure entropy are merged by a new method which is based on the Tsallis nonextensive statistical mechanics.
In the proposed structure entropy, the relationship between the basic factor and the structure entropy is described by the nonextensive additivity. The betweenness is used as a nonevtensive parameter to describe the nonextensive additivity.
The results of our research show that the new structure entropy is more reasonable and more effective to illuminate the structure property of the complex networks, especially for the networks which have a special structure.

\section{Acknowledgment}
The work is partially supported by National Natural Science Foundation of China (Grant No. 61174022), Specialized Research Fund for the Doctoral Program of Higher Education (Grant No. 20131102130002), R$\&$D Program of China (2012BAH07B01), National High Technology Research and Development Program of China (863 Program) (Grant No. 2013AA013801), the open funding project of State Key Laboratory of Virtual Reality Technology and Systems, Beihang University (Grant No.BUAA-VR-14KF-02). Fundamental Research Funds for the Central Universities No. XDJK2015D009.
\bibliographystyle{elsarticle-num}
\bibliography{zqreference}







\end{document}